\documentclass[a4paper]{article}
\usepackage{amsfonts}

\usepackage{epsfig}
\usepackage{amssymb}
\usepackage{amsmath}
\usepackage{amsthm}
\usepackage{latexsym}
\usepackage{graphicx}
\def\be{\begin{equation}}
\def\ee{\end{equation}}
\def\bc{\begin{center}}
\def\ec{\end{center}}

\title{Multitasking network with fast noise.}
\author{Elena Agliari  \footnote{
    Universit\`a di Parma, Dipartimento di Fisica and INFN Gruppo di Parma, Italy}, Adriano Barra  \footnote{
    Sapienza Universit\`a di Roma, Dipartimento di Fisica and GNFM Gruppo di Roma,  Italy}, Andrea Galluzzi
    \footnote{Sapienza Universit\`a di Roma, Dipartimento di Matematica, Italy}, Marco Isopi  \footnote{
    Sapienza Universit\`a di Roma, Dipartimento di Matematica, Italy}}

\begin{document}
\maketitle

\begin{abstract}
We consider the multitasking associative network in the low-storage limit and we study its phase diagram with respect to the noise level $T$ and the degree $d$ of dilution in pattern entries. We find that the system is characterized by a rich variety of stable states, among which pure states, parallel retrieval states, hierarchically organized states and symmetric mixtures (remarkably, both even and odd), whose complexity increases as the number of patterns $P$ grows.
The analysis is performed both analytically and numerically: Exploiting techniques based on partial differential equations, allows us to get the self-consistencies for the order parameters. Such self-consistence equations are then solved and the solutions are further checked through stability theory to catalog their organizations into the phase diagram, which is completely outlined at the end.
This is a further step toward the understanding of spontaneous parallel processing in associative networks.
\end{abstract}

\section{Introduction} \label{sec:intro}
The paradigm, introduced almost three decades ago by Amit, Gutfreund and Sompolinsky \cite{Amit-2003book,Amit-PRL1985}, of analyzing neural networks through techniques stemmed from statistical mechanics of disordered systems (mainly the celebrated Replica Trick \cite{Mezard-1987book} for the Hopfield model \cite{Hopfield-1982PNAS}) has been so prolific that its applications went far beyond Artificial Intelligence and Robotics, overlapping Statistical Inference \cite{Cheng-NeuralNetworks1994}, System Biology \cite{Floreano-2000book}, Financial Market planning \cite{Trippi-1992book}, Theoretical Immunology \cite{Agliari-JTB2011} and much more.
\newline
As a result, research in this field is under continuous development, ranging from the diverse applications outlined above, to an ever deeper understanding of the core-theory behind. For the sake of reaching results closer to experimental neuroscience outcomes, scientists involved in the field tried to bypass the rather crude mean field description of a fully connected network of interacting neurons, embedding them in diluted topologies as Erd\"{o}s-R\'enyi graphs \cite{Sompolinsky-1986PRA}, small-worlds \cite{Coolen-2004JPA} or even finitely connected graphs \cite{Coolen-2003JPA}. The main point was showing robustness of the mean-field paradigm even in these diluted, and in some sense `closer to biology", versions and this was indeed successfully achieved (with the exception of too extreme degrees of dilution, where the associative capacities of the network trivially break down).

Recently, a mapping between Hopfield networks  and Boltzmann machines \cite{Barra-2010JSP,Barra-2012NN} allowed the introduction of dilution into associative networks from a different perspective with respect to standard link removal \`{a} la Sompolinsky \cite{Sompolinsky-1986PRA} or  \`{a} la Coolen \cite{Coolen-2004JPA,Coolen-2003JPA}. In fact, while in their papers these authors perform dilution directly on the Hopfield network, through the equivalence with Boltzmann machine, one may perform link dilution on the Boltzmann machine and then map back the latter into the associative Hopfield-like network checking for its emerging properties \cite{Agliari-2012PRL}. Remarkably, the resulting model still works as an associative performer, as the Hebbian structure is preserved, but its capabilities are quite different from the standard scenario. In particular, the resulting associative network may still be fully-connected but the stored patterns of information display entries  which, beyond coding information through digital values $\pm 1$, can also be blank  \cite{Agliari-2012PRL,Agliari-2012NN}. In fact, any missing link in the bipartite Boltzmann machine corresponds to a blank entry in the related pattern of the associative network.
\\
Now, while standard (i.e., performed directly on the Hopfield network) dilution does not change qualitatively the system performances, the behavior of the system resulting from hidden (i.e., performed on the underlying Boltzmann machine) dilution becomes `multitasking" because retrieval of a single pattern, say $\xi^1$, does not exhaust the whole neurons, and the ones coupled with the $\xi^1$ blank entries are free to align with $\xi^2$, whose entries will partially be blank as well, hence eliciting, in turn, the retrieval of $\xi^3$ and so on up to a parallel logarithmic (with respect to the volume of the network $N$) load of all the stored patterns.
\newline
As a consequence, by tuning the degree of dilution in the hidden Boltzmann network and the level of noise in the directed network, the system exhibits a very rich phase diagram, whose investigation is the subject of the present work.

The paper is organized as follows. In section $2$, we review the multitasking networks introduced in \cite{Agliari-2012PRL} highlighting their main features and providing a rigorous solution for their thermodynamics  through a novel technique based on mapping the statistical mechanical problem into a diffusion problem and then solving the latter through standard partial differential equation methods. In section $3$ solutions obtained in the previous section are investigated. In particular, we discuss the emergence of spurious states for these multitasking networks. Then, in section $4$ we describe the analytical technique used to study the stability of the retrieval states, which are found to be solutions of the system. Exact analytical investigations and numerical results are presented in section $5$ and a very rich phase diagram, where different emergent behaviors in the organization of the neural states are proved. Finally, section $6$ is devoted to a summary and a discussion of the results which are successfully checked against Monte Carlo simulations.

\section{The multitasking associative network}\label{sec:model}
In the conventional Hopfield model (see, e.g., \cite{Amit-2003book,Coolen-2005book}), one considers a network of $N$ neurons, where each neuron $\sigma_i$ can take two states, namely, $\sigma_i = +1$ (firing) and $\sigma_i = -1$ (quiescent). Neuronal states are given by the set of variables $\pmb{\sigma} = (\sigma_1, ... , \sigma_N )$. Each neuron is located on a node of a complete graph and the synaptic connection between two arbitrary neurons, say, $\sigma_i$ and $\sigma_j$, is defined by the following Hebb rule \cite{Amit-2003book}:
\begin{equation} \label{eq:hebb}
J_{ij} = \frac{1}{N} \sum_{\mu=1}^P \xi_i^{\mu} \xi_j^{\mu},
\end{equation}
where $\pmb{\xi}^{\mu}=(\xi_1,...,\xi_N)$ denotes the set of memorized patterns, each specified by a label $\mu=1,...,P$.
The entries are dichotomic, i.e., $\xi_i^{\mu} \in \{ +1, -1\}$, chosen randomly and independently with equal probability, namely, for any $i$ and $\mu$,
\begin{equation} \label{eq:pattern_equi}
P(\xi_i^{\mu}) = \frac{1}{2} (\delta_{\xi_i^{\mu}-1} +  \delta_{\xi_i^{\mu}+1}),
\end{equation}
where the Kronecker delta $\delta_{x}$ equals $1$ iff $x=0$, otherwise it is zero.
Patterns are assumed as quenched, that is, the performance of the network is analyzed keeping the synaptic values fixed.

The Hamiltonian describing this system is
\begin{equation} \label{eq:hopfield}
H_N (\pmb{\sigma , \xi}) = - \sum_{i=1}^N \sum_{i > j=1}^N J_{ij} \sigma_i \sigma_j =  - \frac{1}{2N} \sum_{\substack{i,j=1 \\ j \neq i}}^{N,N}  \sum_{\mu=1}^P \xi_i^{\mu} \xi_j^{\mu} \sigma_i \sigma_j,
\end{equation}
so that the signal (i.e. the field) acting on neuron $i$ is
\begin{equation}
h_i (\pmb{\sigma , \xi})= \sum_{\substack{j=1 \\ j \neq i}}^N J_{ij} \sigma_j.
\end{equation}
The evolution of the system is ruled by a stochastic dynamics, according to which the probability that the activity of a neuron $i$ assumes the value $\sigma_i$ is
\begin{equation} \label{eq:glauber}
P(\sigma_i; \pmb{\sigma , \xi}, \beta ) = \frac{1}{2} [1 + \tanh (\beta h_i \sigma_i)],
\end{equation}
where $\beta$ tunes the level of noise such that for $\beta \to 0$ the system behaves completely randomly, while for $\beta \to \infty$ it becomes noiseless and deterministic; notice that the noiseless limit of Eq.~(\ref{eq:glauber}) is $\sigma_i(t+1) = \textrm{sign} \left [ h_i(t) \right]$.

The main feature of the model described by Eqs.~(\ref{eq:hopfield}) and (\ref{eq:glauber}) is its ability to work as an associative memory. More precisely, the patterns are said to be memorized if each of the network configurations $\sigma_i = \xi_i^{\mu}$ for $i = 1, ..., N$, for everyone of the $P$ patterns labeled by $\mu$, is a fixed point of the dynamics. Introducing the overlap $m^{\mu}$ between the state of neurons $\pmb{\sigma}$ and one of the patterns $\pmb{\xi}^{\mu}$, as
\begin{equation} \label{eq:overlap}
m^{\mu} = \frac{1}{N} (\pmb{\sigma \cdot \xi^{\mu}}) = \frac{1}{N} \sum_{i=1}^N \sigma_i \xi_i^{\mu},
\end{equation}
such a pattern is said to be retrieved if, in the thermodynamic limit, $m^{\mu} = \mathcal{O}(1)$.
Given the definition (\ref{eq:overlap}), the Hamiltonian (\ref{eq:hopfield}) can also be written as
\begin{equation} \label{eq:hopfield2}
H_N (\pmb{\sigma , \xi}) = - N \sum_{\mu=1}^P (m^{\mu})^2 + P = -  N \pmb{m}^2 + P.
\end{equation}

The analytical investigation of the system is usually carried out in the thermodynamic limit $N \rightarrow \infty$, consistently with the fact that real networks are comprised of a very large number of neurons. Dealing with this limit, it  is convenient to specify the relative number of stored patterns, namely $P/N$ and to define the ratio $\alpha = \lim_{N \rightarrow \infty} P/N$. The case $\alpha=0$, corresponding to a number $P$ of stored patterns scaling sub-linearly with respect to the amount of performing neurons $N$, is often referred to as ``low storage''. Conversely, the case of finite $\alpha$ is often referred to as ``high storage''.
In particular, in the former case ($\alpha=0$), the overall behavior of the standard Hopfield model is ruled only by the noise $T \equiv 1/\beta$ and the so-called pure-state ansatz
\begin{equation}
\pmb{m} = (m,0,...,0),
\end{equation}
always corresponds to a stable solution for $T<1$; the order in the entries is purely conventional and here we assume that the first pattern is the one stimulated.

Let us now move on and generalize the system described above in order to account for the existence of blank entries in the patterns $\xi$'s.
More precisely, we replace
 Eq.~(\ref{eq:pattern_equi}) by
\begin{equation} \label{eq:pattern_blank}
P(\xi_i^{\mu}) = \frac{1-d}{2} \delta_{\xi_i^{\mu}-1} +  \frac{1-d}{2} \delta_{\xi_i^{\mu} +1} + d \delta_{\xi_i^{\mu}},
\end{equation}
where $d$ encodes the degree of ``dilution'' in pattern entries.
Patterns are still assumed as quenched and, of course,  the definitions of the Hamiltonian (\ref{eq:hopfield}) and of the overlaps (\ref{eq:overlap}), with the dynamics provided by (\ref{eq:glauber}) still hold.

As discussed in \cite{Agliari-2012PRL,Agliari-2012NN,Agliari-sub}, this kind of extension has strong biological motivations and also yields highly non-trivial thermodynamic outcomes.
In fact, the distribution in Eq.~(\ref{eq:pattern_equi}) necessarily implies that
the retrieval of a unique pattern does employ all the available neurons, so that no resources are left for further tasks.
Conversely, with Eq. (\ref{eq:pattern_blank}) the retrieval of one pattern still allows available neurons (i.e., those corresponding to the blank entries of the retrieved pattern), which can be used to recall other patterns up to the exhaustion of all neurons. The resulting network is therefore able to process several patterns simultaneously.

In particular, in the low-storage regime, it was shown both analytically (via density of states analysis) and numerically (via Monte Carlo simulations) \cite{Agliari-2012PRL,Agliari-sub}, that the system
evolves toward an equilibrium state where several patterns are simultaneoussly retrieved. In the noiseless limit $T=0$ and for $d$ not too large, the equilibrium state is characterized by a hierarchical overlap
\begin{equation}\label{eq:ansatz}
\pmb{m}=(1-d)(1,d,d^2,...,0),
\end{equation}
hereafter referred to as ``parallel ansatz''. On the other hand, in the presence of noise or for large degrees of dilution in pattern entries, this state ceases to be a stable solution for the system and different states, possibly spurious, emerge. Aim of this work is to highlight the equilibrium states of this system as a function of the parameters $d$ and $T$, and finally build a phase diagram; to this task we develop, at first, a rigorous mathematical treatment for calculating the free energy of the model and then obtain the self-consistencies constraining the phase-diagram; then, we solve these equations both numerically and with a stability analysis. In this way we are able to draw the phase diagram, whose peculiarities lie in the stability of both even and odd mixture of spurious states (in proper regions of the parameters) and the formation of parallel spurious state. Both these results generalize the standard counterpart of classical Hopfield networks.
\newline
Findings are double-checked through Monte Carlo runs that are in excellent agreement with the picture we obtained.

\subsection{Statistical mechanics analysis through Fourier technique} \label{sec:LS}
We solve the general model described by the Hamiltonian (\ref{eq:hopfield}), with patterns diluted according to (\ref{eq:pattern_blank}),
in the low storage regime $P \sim \log N$, such that the limit $\alpha=\lim_{N \to \infty} P/N =0$ holds\footnote{Results outlined within this scaling can be extended with little effort to the whole region $P \sim N^{\gamma}$, with $\gamma<1$, such that the constraint $\alpha=0$ is preserved, as realized in the Willshaw model \cite{Willshaw-1976PRSL} concerning neural sparse coding.}.
Due to the formal analogy with statistical-mechanics models for magnetic systems \cite{Amit-2003book}, in the following neurons will be also referred to as spins.

As standard in disordered statistical mechanics, we introduce three types of average for an average observable $o(\pmb{\sigma}, \pmb{\xi})$: $i.$ the Boltzmann average $\omega(o) = \sum_{\sigma} o(\pmb{\sigma}, \pmb{\xi}) \exp[-\beta \mathcal{H}(\pmb{\sigma}; \pmb{\xi})]/Z_{N,P}(\beta,d)$,  where
$$
Z_{N,P}(\beta,d) = \sum_{\{\sigma\}}\exp\left[ - \beta H_N(\pmb{\sigma}, \pmb{\xi}) \right]
$$
is called ``partition function", $ii.$ the average $\mathbb{E}$ performed over the quenched disordered couplings $\xi$, $iii.$ the global expectation $\mathbb{E}\omega(o)$ defined by the brackets $\langle o \rangle_{\xi}$.
\newline
Given these definitions, for the average energy of the system $E$ we can write $E \equiv \lim_{N \to \infty} (\langle H_N(\pmb{\sigma}, \pmb{\xi}) \rangle/N)$.
\newline
Also, we are interested in finding an explicit expression for the order parameters of the model, namely the averaged  P Mattis magnetizations
\begin{equation}
\langle m^\mu \rangle =  \lim_{N \to \infty}\mathbb{E} \omega(\frac{1}{N}\sum_j^N \xi_j^\mu \sigma_j).
\end{equation}
To this task we need to introduce the statistical pressure
$$\alpha(\beta,d)=\lim_{N \to \infty}\frac{1}{N} \ln(Z_{N,P}(\beta,d)),$$
which is immediately  related to the free energy per site  $f(\beta,d)$ by the relation $f(\beta,d)= -\alpha(\beta,d)/\beta$ because, by  maximizing  $\alpha(\beta,d)$ with respect to the P magnetizations $\langle m^{\mu} \rangle$, we  get exactly  the self consistence equations for these order parameters, whose solutions will give us a picture of the phase diagram.

In the past decades, scientists involved in disordered statistical mechanics investigations, even  beyond Artificial Intelligence, paved several strands for solving this kind of problems, and nowadays a plethora of techniques is available. We extend early ideas of Guerra \cite{Barra-2010JSM}, on the line developed in \cite{Genovese-2009JMP}, consisting in modeling disordered statistical mechanics through dynamical system theory and in particular, here, we are going to proceed as follows:
\newline
Our statistical-mechanics problem is mapped into a diffusive problem embedded in a $P$-dimensional space and with given, known, boundaries. We solve the diffusive problem via standard Green-propagator technique, and then we will map back the obtained solutions in terms of their original statistical mechanics meaning.
\newline
To this task, let us introduce and consider a generalized Boltzmann factor $B_N(\textbf{x},t)$ depending on $P+1$ parameters $\textbf{x}, t$ (which we think of as \emph{generalized P dimensional Euclidean space} and \emph{time})
\begin{equation}
B_N(\mathbf{x},t;\pmb{\xi},\pmb{\sigma})=\exp\bigg(\frac{t}{2N}\sum_{i\neq j}^N\sigma_i \sigma_j \sum_{\mu}^P\xi_i^\mu \xi_j^\mu+\sum_\mu^P x_\mu \sum_j^N \xi_j^\mu \sigma_j \bigg),
\end{equation}
and the generalized statistical pressure
\begin{equation}
\alpha_N(\textbf{x},t)=\frac{1}{N}\ln\left[\sum_{\mathbf{\{\sigma\}}}B_N(\mathbf{x},t;\pmb{\xi},\pmb{\sigma} )  \right].
\end{equation}
Notice that, for proper values of $\textbf{x}, t$, namely  $\textbf{x}=0$ and  $t=\beta$, classical statistical mechanics is recovered as
\begin{equation}
\alpha(\beta)=\lim_{N \to \infty}\alpha_N(\mathbf{x}=0,t=\beta)=\lim_{N \to \infty}\frac{1}{N}\ln\left [\sum_{\mathbf{\{\sigma\}}}B_N(\mathbf{x}=0,t=\beta;\pmb{\xi},\pmb{\sigma} ) \right].
\end{equation}
In the same way, the average $\langle \cdot \rangle_{(\mathbf{x},t)}$ will be denoted by $\langle \cdot \rangle$, wherever evaluated in the sense of statistical mechanics, namely
\begin{equation}
\langle o \rangle_{(\mathbf{x},t)}=\frac{\sum_{\mathbf{\{\sigma\}}} o(\pmb{\sigma},\pmb{\xi})B_N(\mathbf{x},t;\pmb{\xi},\pmb{\sigma} )  }{\sum_{\{\sigma\}}B_N(\mathbf{x},t;\pmb{\xi},\pmb{\sigma} )},
\end{equation}
\begin{equation}
\langle o \rangle=\frac{\sum_{\mathbf{\{\sigma\}}} o(\pmb{\sigma},\pmb{\xi})\exp[-\beta H(\pmb{\sigma},\pmb{\xi})] }{\sum_{\mathbf{\{\sigma\}}}\exp[-\beta H(\pmb{\sigma},\pmb{\xi})]}=\langle o \rangle_{(\mathbf{x}=0,t=\beta)}.
\end{equation}
It is immediate to see that the following equations hold:
\begin{equation}
  \begin{array}{ll}
     \partial_t \alpha_N(\mathbf{x},t)=\frac{1}{2}\sum_\mu \langle m_\mu^2\rangle_{(\mathbf{x},t)},  \\
     \partial_{x_\mu} \alpha_N(\mathbf{x},t)= \langle m_\mu\rangle_{(\mathbf{x},t)},
  \end{array}
\end{equation}
and, defining a vector $\Gamma_N(\mathbf{x},t)$ of elements $\Gamma^\mu_N(\mathbf{x},t) \equiv -\partial_{x_\mu}\alpha_N(\mathbf{x},t)$, by construction $ \Gamma^\mu_N(\mathbf{x},t)$ obeys the following equation:
\begin{equation}
\partial_t \Gamma^\mu_N(\mathbf{x},t) +\sum_{\nu=1}^P \Gamma^\nu_N(\mathbf{x},t) [ \partial_{x_\nu} \Gamma^\mu_N(\mathbf{x},t) ]=\frac{1}{2N}\sum_{\nu=1}^P \partial^2_{x_\nu^2}\Gamma^\mu_N(\mathbf{x},t),
\end{equation}
which happens to be in the form of a Burgers' equation for the vector $\Gamma_N(\mathbf{x},t)$ with a kinematic viscosity $(2N)^{-1}$.
As it is well-known, the Burger equation can be mapped into a $P$-dimensional diffusive problem using the Cole-Hopf transformation \cite{Genovese-2009JMP} as follow:
\begin{equation}
\psi_N(\mathbf{x},t)=\exp\left[-N\int dx_\mu \Gamma^\mu_N(\mathbf{x},t)\right]=\exp[N\alpha_N(\textbf{x},t)],
\end{equation}
and its $t$ and $x$ streaming read off as
\begin{equation}
  \begin{array}{ll}\label{eq:derivate}
     \partial_t \psi_N(\mathbf{x},t)=N(\partial_t \alpha_N(\textbf{x},t) ) \psi(\mathbf{x},t), \\
     \partial_{x_\mu} \psi_N(\mathbf{x},t)=N(\partial_{x_\mu} \alpha_N(\textbf{x},t) ) \psi(\mathbf{x},t),
  \end{array}
\end{equation}
in such a way that
\begin{equation}\label{eq:derivate2}
      \partial^2_{x_\mu x_\nu} \psi_N(\mathbf{x},t)=N \psi_N(\mathbf{x},t) \left\{ \partial^2_{x_\mu x_\nu}\alpha_N(\mathbf{x},t)+N [ \partial_{x_\mu }\alpha_N(\mathbf{x},t)] [ \partial_{x_\nu }\alpha_N(\mathbf{x},t)]   \right\}.
\end{equation}
Now, from equations \eqref{eq:derivate}, \eqref{eq:derivate2} we get
\begin{equation}
     \partial_t \psi_N(\mathbf{x},t)-\frac{1}{2N}\sum_\mu \left[ \partial^2_{x_\mu^2}\psi_N(\mathbf{x},t) \right]=0.
\end{equation}
Therefore, we established a reformulation of  the problem of calculating the thermodynamic potential $\alpha(\beta, d)$ over the equilibrium configuration of the order parameters for an attractors network model in terms of a diffusion equation for the function $\psi_N(\mathbf{x},t)$, namely the Cole-Hopf transform of the Mattis magnetizations, with a diffusion coefficient $D=(2N)^{-1}$, that is
$$
     \partial_t \psi_N(\mathbf{x},t)-D \nabla^2 \psi_N(\mathbf{x},t) =0,
$$
\begin{equation}\label{eq:Cauchy}
      \psi_N(\mathbf{x},0)=\sum_{\{\sigma\}}\exp\bigg(\sum_\mu x_\mu \sum_j \xi_j^\mu \sigma_j\bigg).
\end{equation}
We solve this Cauchy problem \eqref{eq:Cauchy} through standard techniques: first,  we map the diffusive equation in the Fourier space, then
we calculate the Green propagator for the homogenous configuration, and finally we will inverse-transform the solution.
\newline
Let us consider the Fourier transform:
\begin{equation}
  \begin{array}{ll}
      \widetilde{\psi}_N(\mathbf{k},t)= \int_{\mathbb{R}^P} d^P x \exp\big(-i\sum_\mu x_\mu k_\mu\big)\psi_N(\mathbf{x},t), \\
     \psi_N(\mathbf{x},t)=\frac{1}{(2 \pi)^P} \int_{\mathbb{R}^P} d^P k \exp\big(i\sum_\mu x_\mu k_\mu\big)\widetilde{\psi}_N(\mathbf{k},t),
  \end{array}
\end{equation}
and the related Green problem:
\begin{equation}
      \partial_t \widetilde{G}(\mathbf{k},t)+D k^2 \widetilde{G}(\mathbf{k},t)=\delta(t),
\end{equation}
where $\widetilde{G}(\textbf{k},t)$ is the Green propagator in the $k$-space, which can be decomposed as
 \begin{equation}
        \widetilde{G}(\textbf{k},t)=\widetilde{G}_R(\textbf{k},t)+\widetilde{G}_S(\mathbf{k},t),
\end{equation}
being $\widetilde{G}_R(\textbf{k},t)$ the general solution of the homogeneous problem and $\widetilde{G}_S(\textbf{k},t)$ a particular solution of the non-homogeneous problem.
Hence, the full solution will be
\begin{equation}
\psi_N(\mathbf{x},t)= \int_{\mathbb{R}^P}d^P x' G_R(\mathbf{x}-\mathbf{x'},t)\psi_N(\mathbf{x'},0),
\end{equation}
where the function $\widetilde{G}_R(\mathbf{k},t)$ fulfills
\begin{equation}
  \begin{array}{ll}
    \partial_t \widetilde{G}_R(\mathbf{k},t) -D k^2 \widetilde{G}_R(\mathbf{k},t)  =0, \\
     \widetilde{G}_R(\mathbf{k},0)=1,
  \end{array}
\end{equation}
hence
\begin{equation}
  \begin{array}{ll}
    \widetilde{G}(\mathbf{k},t)=\exp(-D k^2 t), \\
    G(\mathbf{x},t)=\frac{1}{(2\sqrt{\pi D t})^P}\exp(\frac{-\mathbf{x}^2}{4D t}).
  \end{array}
\end{equation}
Therefore, we get
\begin{equation}
\psi_N(\mathbf{x},t)= \left(\frac{N}{2\pi t} \right)^{\frac{P}{2}}\int (\prod_\mu dx'_\mu) \exp \left[ -N\Phi(\mathbf{x'},\mathbf{x},t) \right] ,
\end{equation}
\begin{equation}
\Phi(\mathbf{x'},\mathbf{x},t)= \frac{\sum_\mu^P (x_\mu-x'_\mu)^2}{2t}-\ln2-\frac{1}{N}\sum_{j=1}^N \ln\left[\cosh \left(\sum_\mu x'_\mu \xi_j^\mu \right)\right]
\end{equation}
and
\begin{equation}
\alpha_N(\mathbf{x},t)= \frac{1}{N}\ln \left[ \psi_N(\mathbf{x},t) \right].
\end{equation}
We can solve now the saddle-point equation
\begin{equation}
\alpha(\mathbf{x},t)=\lim_{N \to \infty}\alpha_N(\mathbf{x},t)= \textrm{Extr}\{\Phi\},
\end{equation}
where we neglected $ \mathcal{O}(N^{-1})$ terms, as we performed the thermodynamic limit. Finally, by replacing $t=\beta$ and $\mathbf{x}=0$ and $x'_\nu=\beta \langle m_\nu \rangle$  (hence the original statistical mechanics framework),  we obtain the following expressions for the statistical pressure
\begin{equation}\label{eq:freeX}
\alpha(\beta)=\frac{\beta}{2}\sum_\mu \langle m_\mu\rangle^2-\ln(2)- \bigg\langle\ln \left[  \cosh \left (\beta \sum_\mu \langle m_\mu\rangle \xi^\mu \right) \right]\bigg\rangle_\xi,
\end{equation}
whose extremization offers immediately the $P$ desired self-consistency equations for all the  $\langle m_{\nu}\rangle$,
\begin{equation} \label{eq:SC}
\langle m_\nu\rangle=\bigg\langle \xi^\nu \tanh \left (\beta \sum_\mu \xi^\mu \langle m_\mu\rangle \right)\bigg\rangle_\xi  \qquad \forall \mu \in [1,P],
\end{equation}
where with the index $\xi$ we emphasized once more that the disorder average over the quenched patterns is performed as well.

Of course, the self-consistence equations (\ref{eq:SC}) recover those obtained in \cite{Agliari-2012PRL,Agliari-sub} via different analytical techniques, where they were also shown to yield to the parallel ansatz (\ref{eq:ansatz}), which, in turn, can be formally written as
\be \label{eq:parallelstate}
\sigma_i=\xi_i^1+\sum_{\nu=2}^P \xi_i^{\nu}\prod_{\mu=1}^{\nu-1}\delta(\xi_i^{\mu}),
\ee
and it will be referred to as $\sigma^{(P)}$.

The parallel ansatz (\ref{eq:ansatz}) can be understood rather intuitively. To fix ideas let us assume zero noise level and that one pattern, say $\mu=1$, is perfectly retrieved. This means that the related average magnetization is $m_1=(1-d)$, while a fraction $d$ of spins is still available and they can arrange to retrieve a further pattern, say $\mu=2$. Again, not all of them can match non-null entries in pattern $\xi^2$ and the related average magnetization is $m_2=d(1-d)$. Proceeding in the same way, for  all spins, we get the parallel state. Notice that, the number $K$ of patterns which are, at least partially, retrieved does not necessarily equal $P$. In fact, due to discreteness, it must be $d^{K-1}(1-d)\leq 1/N$, namely at least one spin must be aligned with $\xi^{K}$, and this implies $K \lesssim \log N$.

Such a hierarchical, {\em parallel}, fashion for alignment, providing an overall energy (see Eq.~\ref{eq:hopfield2})
\be
E^{\textrm{(P)}} = - N  \sum_{k=1}^{P} [(1-d)d^{k-1}]^2 + P = - N \frac{(1-d^{2P})(1-d)}{1+d} + P,
\ee
is more optimal than a {\em uniform} alignment of spins amongst the available patterns, as this case would yield $m_k=(1-d)/P$ for any $k$ and an overall energy
\be
E^{\textrm{(U)}} = - N  \sum_{k=1}^{P} \left( \frac{1-d}{P} \right)^2+ P = - \frac{(1-d)^2 N}{P} + P,
\ee
being $(1-d^{2+2P}) > (1-d^2)/P$.
\newline
On the other hand, as we will see in Sec.~\ref{sec:failure}, when $d > d_{c} \approx 1/2$, the state (\ref{eq:ansatz}) is no longer stable and spurious states do emerge.

Before proceeding, it is worth stressing that, although the parallel state (\ref{eq:ansatz}) displays non-zero overlap with several patterns, it is deeply different, and must not be confused with, a spurious state in standard Hopfield networks.
In fact, in the former case, at least one pattern is completely retrieved, while in spurious states, the overlap with each memory pattern involved is only partial.
\newline
Moreover, in standard Hopfield networks, spurious states are somehow undesirable because they provide corrupted information with respect to the best retrieval achievable where one, and only one, pattern is exactly retrieved. Conversely, in our model, the retrieval of more-than-one pattern is unavoidable (for finite $d$ and $\beta \to \infty$) and the quality of retrieval may be excellent (perfect) in the case of patterns poorly (not) overlapping.
\newline
Finally, and most importantly, for $\beta \to \infty$ and in a wide region of dilution, the parallel state $\sigma^{(P)}$ corresponds to a global minimum for the energy. This is not the case for an arbitrary mixture of states.


\section{The emergence of spurious states} \label{sec:spuri}

In Sec.~\ref{sec:LS}, we explained why we expect the parallel state (\ref{eq:parallelstate}) to occur, exploiting the fact that each pattern tends to align as many spins among those still available. Actually, this intuitive approach yealds the correct picture for $T=0$ (no fast noise) and not-too-large $d$, while when either $T$ or the degree of dilution are large enough, the system can relax to a state where only one pattern is retrieved or falls into a spurious state where several patterns are partially retrieved, but none exactly. These states are discussed in the following subsections and in Sec.~\ref{sec:stability} the analysis will be made quantitative.

\subsection{The failure of parallel retrieval} \label{sec:failure}
Let us start from the noiseless case and consider the state (\ref{eq:parallelstate}) corresponding to the parallel ansatz (\ref{eq:ansatz}): we notice that, on average, there exists a fraction $2[(1-d)/2]^P$ of spins $\sigma_i$ corresponding to the entries $\xi_i^1 = 1, \xi_i^k = -1, \forall k \in [1,P]$ (and analogously for the ``gauged'' case $\xi_i^1 = - 1, \xi_i^k = +1)$ and expected to be aligned with the first entry $\xi_i^1$, in such a way that the overall field insisting on each of them is $h_i = m_1 - m_2 - m_3 - .... - m_P$. Of course, such spins are the most unstable, and, at zero noise level, they flip whenever $h_i$ happens to be negative, that is, when $m_1 < \sum_{k=2}^P m_k$. Exploiting the ansatz $m_k=d^{k-1}(1-d)$, this can be written as
\begin{equation} \label{eq:campo}
h_i = (1-d) \left [ 1 - \frac{d-d^P}{1-d} \right] = 1-2d +d^P,
\end{equation}
which becomes negative for a value of dilution $d_{c}(P)$, which converges exponentially from above to $1/2$ as $P$ gets large.
From this point onwards, the first pattern is no longer completely retrieved and the system fails to parallel retrieve (according to the definition in Eq.~\ref{eq:parallelstate}).
Therefore, when $d \geq d_{c}(P)$, genuine spurious states emerge and the system relaxes to states which correspond to mixture of $p \leq P$ patterns, but none of them is completely retrieved (at least up to extreme values of dilution).
As we will see in Sec.~\ref{sec:parallel}, the transition at $d_c(P)$ is first order.

Moreover, from Eq.~\ref{eq:campo} we find that the case $P=2$ has no solution in the range $d \in [0,1]$, meaning that the parallel-retrieval state is always a stable solution in the zero noise limit; on the other hand, $d_{c}(3) \approx 0.62$, $d_{c}(4) \approx 0.54$ and so on.


Such phenomenology concerns relatively large degrees of dilution, yet, the presence of noise can also destabilize the true parallel-retrieval state (\ref{eq:ansatz}) in the regime of small degrees of dilution. In fact, we expect that the spins aligned according to the $k$-th pattern associated to a magnetization $m_k=d^{k-1}(1-d)$ will loose stability at noise levels $T>d^{k-1}(1-d)$. In particular, at $T>d(1-d)$, only one pattern will be retrieved and the pure state is somehow recovered. As we will see in Sec.~\ref{sec:parallel}, such estimates are correct for small $d$.

\subsection{Symmetric mixtures}
Typical spurious states emerging in standard associative networks are the so-called symmetric mixtures of $p\leq P$ states, which can be described as
\begin{equation} \label{eq:symmstate}
\sigma_i = \textrm{sign}\left (\sum_{\mu =1}^{p} \xi_i^{\mu} \right),
\end{equation}
and it will be referred to as $\sigma^{(S)}$.
We anticipate that the symmetric mixture turns out to emerge also in the diluted model under investigation.
\newline
Now, in the standard Hopfield model, odd mixtures of $p$ patterns, are metastable, i.e. their energies are higher than those of the pure patterns, and, moreover, the smaller $p$ and the more energetically favorable the mixture.
On the other hand, even mixtures of $p$ patterns are unstable (they are saddle-points of the energy).
%
%
The instability of even mixtures is often associated to the fact that, for a macroscopic fraction of spins, $\sigma^{(S)}$ is not defined due to the ambiguity of the $\textrm{sign}$. For instance, when $p=2$, $\sum_{\mu =1}^{p} \xi_i^{\mu}$ occurs to be null for half of the spins and the related values are defined stochastically according to the distribution
\be
P(\sigma_i) = \frac{1}{2}(\delta_{\sigma_i+1} + \delta_{\sigma_i+1}).
\ee

However, as we will show in Sec.~\ref{ssec:simm}, this is not the case for this diluted model as it displays wide regions in the parameter space $(d,T)$ where even and/or odd symmetric mixtures are stable.

%

\subsection{A ``hybrid'' spurious state}\label{sec:ibrido}
As we will see in Sec.~\ref{ssec:simm}, the symmetric mixture $\sigma^{(S)}$ can become unstable and relax to a different spurious state which is a ``hybrid'' state between the symmetric mixture $\sigma^{(S)}$ and the parallel state $\sigma^{(P)}$.

To begin and fix ideas, let us set $P=3$ and start from the state $\sigma_i=\textrm{sign}(\xi^1_i+\xi^2_i+\xi^3_i)$. In the presence of dilution the argument $\xi^1_i+\xi^2_i+\xi^3_i$ can be zero and in that situation one can adopt the following hierarchical rule: take $\sigma_i = \xi_i^{1}$ provided that $\xi_i^{1} \neq 0$; otherwise, if $\xi_i^{1} = 0$, then take $\sigma_i = \xi_i^{2}$ provided that $\xi_i^{2} \neq 0$; otherwise, if also $\xi_i^{2} = 0$, then take $\sigma_i = \xi_i^{3}$ provided that $\xi_i^{3} \neq 0$; otherwise, if also  $\xi_i^{3} \neq 0$, then put $\sigma_i=\pm1$ with probability $1/2$. In this way we can built a state, generally defined for any $P$, and, being $\Xi= \sum_{\mu} \xi_i^{\mu}$, it can written as
\begin{equation} \label{eq:IP}
{
\sigma_i= (1 - \delta_{\Xi,0})  \textrm{sign}(\Xi) +  \delta_{\Xi,0}[   \xi_i^{1} + \delta_{\xi_i^{1},0} \xi_i^{2} + \delta_{\xi_i^{1},0}\delta_{\xi_i^{2},0} \xi_i^{3} + ...]},
\end{equation}
which will be referred to as $\sigma^{(H)}$.

The related average Mattis magnetizations can be calculated as the sum of one contribution $m_0$ (the same for any $\mu$) deriving from the spins corresponding to non ambiguous sign function (i.e., $\Xi \neq 0$), and another contribution accounting for hierarchical corrections (i.e., $\Xi =0$).
Let us focus on the first term:
\begin{eqnarray}
m_0 &=& \langle \xi^{\mu} \textrm{sign}(\Xi) \rangle_{\xi} = \frac{1-d}{2}\left \langle \textrm{sign}(1 + \sum_{\nu \neq \mu}^P \xi^{\mu}) -  \textrm{sign}(- 1 + \sum_{\nu \neq \mu}^P \xi^{\mu}) \right \rangle_{\xi} \\
&=& (1-d) \left [\mathcal{P}(\sum_{\nu \neq \mu}^P \xi^{\nu} < 1) - \mathcal{P}(\sum_{\nu \neq \mu}^P \xi^{\nu} > 1) \right ],
\end{eqnarray}
where, in the last step, we exploited the implicit symmetry in pattern entries and $\mathcal{P}(\sum_{\nu \neq \mu}^P \xi^{\nu} \gtrless 1)$ represents the probability that the specified inequality is verified over the distribution (\ref{eq:pattern_blank}). The latter quantity can also be looked at as the probability for a symmetric random walk with holding probability $d$ to be at distance $\gtrless 1$ from its origin after a time span $P-1$. Hence, we get
\begin{equation}
m_0 = (1-d) [\mathcal{P}(0 \rightarrow 0, P-1) + \mathcal{P}(0 \rightarrow 1, P-1)],
\end{equation}
where $\mathcal{P}(x_0 \rightarrow x, t)$ is the probability for a symmetric random walk with stopping probability $d$ to move from site $x_0$ to site $x$ in $t$ steps, namely
\begin{equation}
\mathcal{P}(x_0 \rightarrow x, t) = \sum_{s=0}^{t-(x-x_0)} \frac{t!}{ s! \left(\frac{t-s-(x-x_0)}{2}\right)!\left( \frac{t-s+(x-x_0)}{2}\right)!} d^s \left( \frac{1-d}{2} \right)^{t-s}.
\end{equation}

The second contribution to the magnetization is $(1-d)  \sum_{k=1}^{P-1} \mathcal{P}(0 \rightarrow 1, P-k)d^{k-1}$.

Finally, by summing the two contributions we find the following expressions for $P=3$
\begin{eqnarray}
m_1 &=& \frac{1}{2} (1 + d - 3 d^2 + d^3),\\
m_2 &=& \frac{1}{2} (1 - d) (1 + d^2), \\
m_3 &=&  \frac{1}{2} (1 - 3d + 5d^2 - 3 d^3 ),
\end{eqnarray}
and for $P=5$
\begin{eqnarray}
m_1 &=& \frac{1}{8} (3 +9 d -42 d^2 + 74d^3 -65 d^4 + 21 d^5), \\
m_2 &=& \frac{1}{8} (1 - d) (3 + 6 d^2 - d^4), \\
m_3 &=& \frac{1}{8} (1 - d) (3 -4d + 18d^2 - 20d^3 + 11 d^4),\\
m_4 &=&  \frac{1}{8} (1 - d) (3 - 4d + 18d^2 -28 d^3 + 19 d^4),\\
m_5 &=&  \frac{1}{8} (1 - d) (3 - 4d +18d^2 -36 d^3 +27 d^4).
\end{eqnarray}
The expressions for arbitrary $P$ can be analogously calculated exactly and some examples are shown in Fig.~$1$.

\begin{figure}[!hd] \label{fig:Ibrido}
\begin{center}
\resizebox{0.90\columnwidth}{!}{\includegraphics{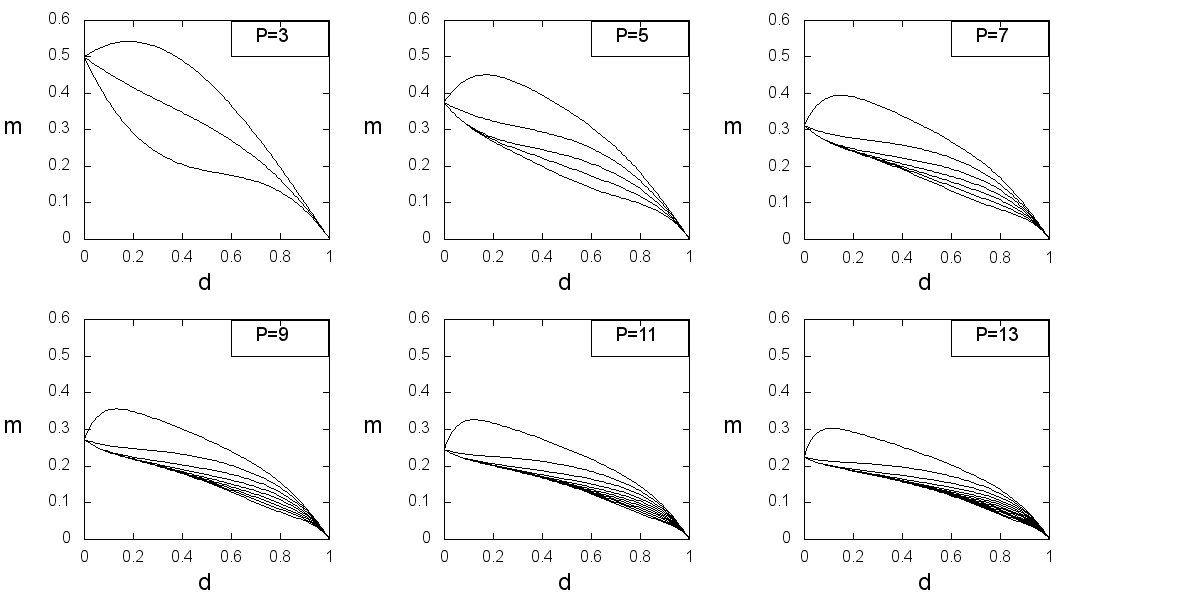}}
\caption{Mattis magnetizations $\mathbf{m}$ versus dilution $d$, according to the analytical expression derived in Sec.~\ref{sec:ibrido}. Each panel refers to a different value of $P$, as specified.}
\end{center}
\end{figure}

We expect $\sigma^{H}$ to become globally stable in the region of very large dilutions ($d>d_{H}(P)$); intuitively, dilution must be large enough to make magnetizations rather close to each other in such a way that the least signalled spins corresponding to $(-,-,...,-,+,+,...,+)$ (overall $(P-1)/2$ negative entries and $(P+1)/2$ positive entries) are stable.
%
This means
$\sum_{i} (1 - \delta_{\Xi,0})  \textrm{sign}(\Xi) \xi_i^{\mu} / N > \sum_{k=1}^{(P-1)/2} \varphi_k (P+1)/(P-k)$, where $\varphi_k=2\sum_l [(1-d)/2]^{2l} d^{P-2l} (P-k)!/[l!(l-1)!(P-k-2l+1)!]$ and $P$ is odd. This condition is fulfilled for values of dilution larger than $d_H(P)$, which converges to $1$ as $P$ gets larger, hence, in order to tackle this limit, dilution must become a function of  the system size $d \to d(N)$. In this case the network itself becomes diluted as well and different techniques are required; this will not be discussed in this paper.


\section{Stability analysis on the organization of the states} \label{sec:stability}

The set of solutions for self-consistent equations (\ref{eq:SC}) describes states whose stability may vary strongly. In fact, provided the network has reached them, in the noiseless limit (of whatever kind) it would persist in those states. However, the equations do not contain any information about whether the solutions will be stable against small perturbations, that is  to say if the system will indeed really thermalize on these states or will fall apart more or less quickly. In order to evaluate their stability we need to check the second derivative of the free-energy \cite{Amit-2003book}.
More precisely, we further need to build up the so called ``stability matrix'' $\mathbf{A}$ with elements
\be  \label{eq:A}
A^{\mu \nu}=\frac{\partial^2 f_\beta(\overrightarrow{m})}{\partial m^\mu \partial m^\nu}.
\ee
Then, we evaluate and diagonalize $\mathbf{A}$ at a point $\mathbf{\tilde{m}}$, representing a particular solution of the self-consistence equations (\ref{eq:SC}), in order to determine whether $\mathbf{\tilde{m}}$ is stable or not. Being $\{ E_{\mu} \}_{\mu=1,...,P}$, the set of related eigenvalues, $\mathbf{\tilde{m}}$ is stable whenever all of them are positive.


%
%

Now, from Eq.~\ref{eq:freeX} and \ref{eq:A}, remembering that $\alpha(\beta,d)=-\beta f(\beta,d)$, we find straightforwardly
\be \label{eq:AA}
A^{\mu \nu} = [1 - \beta (1-d)] \delta^{\mu \nu} + \beta Q^{\mu \nu},
\ee
where
\be\label{eq:BB}
Q^{\mu \nu} = \langle \xi^{\mu} \xi^{\nu} \tanh^2(\beta \overrightarrow{\xi}\cdot\overrightarrow{m}) \rangle_{\xi}.
\ee

Of course when $d=0$ we recover $A^{\mu \nu} = (1- \beta) \delta^{\mu \nu} +  \langle \xi^{\mu} \xi^{\nu} \tanh^2(\beta \overrightarrow{\xi}\cdot\overrightarrow{m}) \rangle_{\xi}$, namely the result known for the standard Hopfield model.



We now consider several states, known to be solutions of self-consistence equations (\ref{eq:SC}) and check their stability. In this way we will find constraints in the region $(T, d)$ where those states are stable and then we will build up the phase diagram.

\subsection{Paramagnetic state} \label{ssec:PM}
Let us start with the paramagnetic state, which is described by
\be
\overrightarrow{m}=\overrightarrow{0};
\ee
this state trivially fulfills Eq.~\ref{eq:SC}.

By replacing this expression in Eq.s ~\ref{eq:AA} and  \ref{eq:BB} we find
\be
A^{\mu \nu}=\delta_{\mu\nu}[1-\beta(1-d)].
\ee
Therefore, in this case, $\mathbf{A}$ is diagonal and its eigenvalues are directly $E_{\mu}= A^{\mu \mu} = 1- \beta(1-d), \forall \nu \in [1,P]$.
We can conclude the paramagnetic state exists and is stable in the region $ 1- \beta(1-d)>0$, that is (remembering that $T=\beta^{-1}$)
\be
\textrm{PM \, stability} \,  \Rightarrow \,  T>1-d.
\ee
This region is highlighted in Fig.~\ref{fig:PM}.
\begin{figure}[!hd]
 \begin{center}
\label{fig:PM}
\resizebox{0.60\columnwidth}{!}{\includegraphics{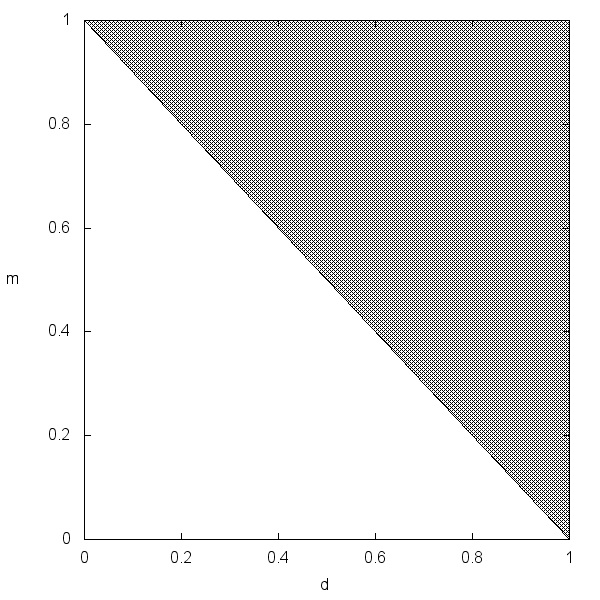}}
\caption{\label{fig:PM} (Color on line) In the parameter space $(T,d)$ we highlighted the region where the paramagnetic state exists and is stable. As proved in Sec.~\ref{ssec:PM}, this region includes points fulfilling $T>1-d$; notice that   this result is independent of $P$.}
\end{center}
\end{figure}

\subsection{Pure state}
Let us now consider the pure state, that is any of the $P$ configurations
\be
\overrightarrow{m}=m(1,\overrightarrow{0}),
\ee
$m$ being the extent of the overlap, which, in general, depends on $d$ and on $T$.
The related self-consistence equations are
\begin{eqnarray}
m^\mu &=&(1-d)\tanh(\beta m^\mu),\\
m^{\nu\neq\mu}&=&0.
\end{eqnarray}
The first equation has solution in the whole half-plane $T>1-d$, and this ensures that, in the same region, the pure-state exists.
In order to check its stability, we calculate the stability matrix finding
\begin{eqnarray}
A^{\mu\nu} &=& 0 \vee \mu\neq\nu\\
A^{\mu\mu} &=& 1-\beta(1-d)[1-\tanh^2(\beta m^\mu)]\\
\label{eq:nunu}
A^{\nu\nu} &=& 1-\beta(1-d)[1-(1-d)\tanh^2(\beta m^\mu)].
\end{eqnarray}
Therefore $\mathbf{A}$ is diagonal and the eigenvalues are  $E_{\mu} = A^{\mu \mu}$ and $E_{\nu} = A^{\nu \nu}$. Notice that these eigenvalues do not depend on $P$ and that $E_{\mu} \geq E_{\nu}$, so that the analysis can be restricted on $E_{\nu}$.
Requiring the positivity for $E_{\nu}$, we get the region in the plane $(T,d)$, where the pure state is stable; such a region is shown in Fig.~\ref{fig:PS}.
We stress that this result is universal with respect to $P$ (in the low-storage regime).
\begin{figure}[!hd]
 \begin{center}
\label{fig:PS}
\resizebox{0.60\columnwidth}{!}{\includegraphics{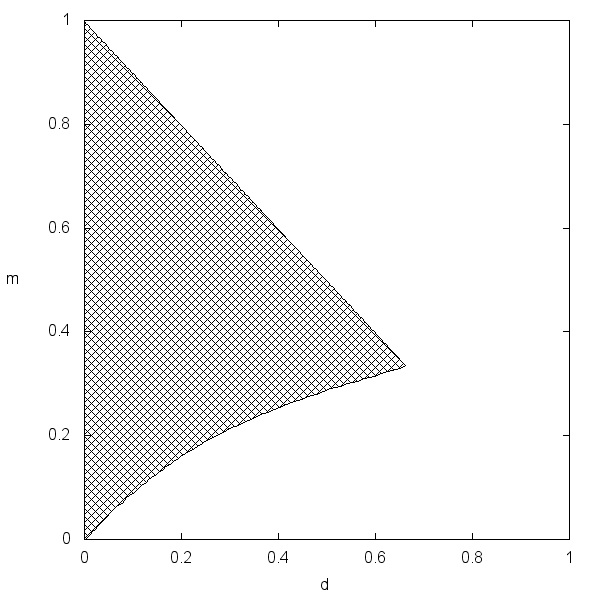}}
\caption{\label{fig:PS} (Color on line) In the parameter space $(T,d)$ we highlighted the region where the pure state exists and is stable. This result was found by numerically solving the self-consistence equation Eq.~\ref{eq:SC} and the inequality $E_{\nu} >0$, where $E_{\nu}$ is the smallest eigenvalues of the stability matrix $\mathbf{A}$ (see Eq.~\ref{eq:nunu}); notice that   this result is independent of $P$.}
\end{center}
\end{figure}

\subsection{Symmetric state}\label{ssec:simm}
A symmetric mixture of states corresponds to configurations leading to
\begin{equation}
\overrightarrow{m}=m(d,T)(1,1,1,...,1,0,...,0),
\end{equation}
where $p\leq P$ order parameters are equivalent and non null, while the remaining $P-p$ are vanishing.

Let us start with the case $p=P=3$, yielding $\overrightarrow{m}=m(d,T)(1,1,1)$.
In this special case the three self-consistence equations collapse on
\begin{eqnarray}
\nonumber
m(d,T) &=& 2 \left ( \frac{1-d}{2} \right )^3 \left[ \tanh^2(3\beta m) + \tanh^2(\beta m) \right] + d \left(\frac{1-d}{2} \right)^2 \tanh^2(2\beta m)\\
&+& 2 \left(\frac{1-d}{2} \right)d^2 \tanh^2(\beta m)
\end{eqnarray}
and the matrix $\mathbf{A}$ reads as
\be
\left(
  \begin{array}{ccc}
    a & b & b \\
    b & a & b \\
    b & b & a \\
  \end{array}
\right)
\ee
$a$ and $b$ being parameters related to $m$, $d$ and $\beta$. More precisely, the eigenvalues of $\textbf{A}$ are $(a+2b,a-b,a-b)$, which can be written as
\be
\nonumber \small
a-b=1-\beta(1-d)+2\beta
\bigg \{
\tanh^2(2\beta m)  d\left(\frac{1-d}{2}\right)^2 + \tanh^2(\beta m) \bigg[ \frac{d^2(1-d)}{2}+
4\left (\frac{1-d}{2}\right)^3 \bigg]\bigg \},
\ee
\begin{eqnarray}
\nonumber
a+2b &=& 1-\beta(1-d)+2\beta \bigg \{ \tanh^2(3\beta m) 3 \left(\frac{1-d}{2} \right)^3
+\tanh^2(\beta m) \bigg[\frac{d^2(1-d)}{2} \\
&+& \left(\frac{1-d}{2}\right)^3\bigg] \bigg \}+ 8d \beta\tanh^2(2\beta m) \left(\frac{1-d}{2}\right)^2.
\end{eqnarray}
The conditions for the existance and the stability of the symmetric, odd mixture with $p=P=3$, yield a system of equations which was solved numerically and the region were such conditions are all fulfilled is shown in Fig.~$4$. 
Notice that the region is actually made up of two disconnected parts, each displaying peculiar features, as explained later.
\newline
This result is robust with respect to $P$, being $P$ odd and $p=P$.

\begin{figure}[!hd] \label{fig:SimmP3}
\begin{center}
\resizebox{0.60\columnwidth}{!}{\includegraphics{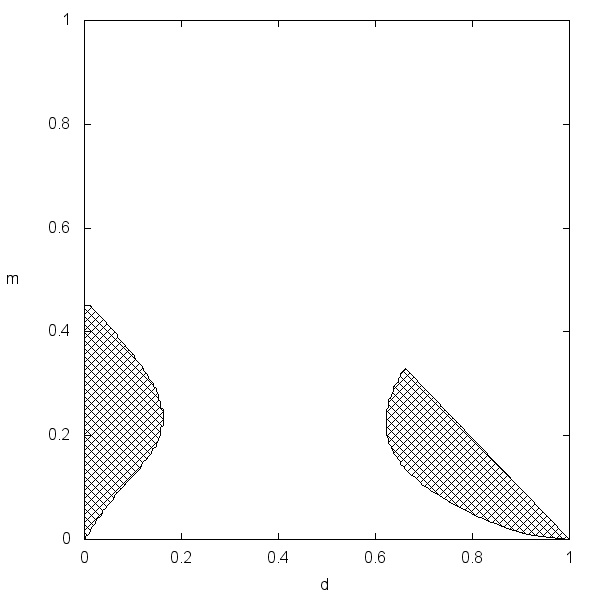}}
\caption{(Color on line)  In the parameter space $(T,d)$ we highlighted the region where the symmetric state $\sigma^{(S)}$, for the special case $p=P=3$, exists and is stable. Notice that two disconnected regions emerge: the one corresponding to lower values of dilution derives from the fact that $p$ is odd, while the one corresponding to larger values of dilution from the fact that $p=P$.}
\end{center}
\end{figure}

We can further generalize the analysis by considering $P>p$, still being $p$ odd. In this case we get the following stability matrix
\be
\left(
  \begin{array}{cccc}
    a & b & b & 0 \\
    b & a & b & 0 \\
    b & b & a & 0 \\
    0 & 0 & 0 & c \\
  \end{array}
\right)
\ee
with eigenvalues $(a-b,a-b,a+2b,c)$, where
\begin{eqnarray}
\nonumber
c&=&1-\beta(1-d)\\
\nonumber
&\times& \Big \{ 1-2 \Big[ \left(\frac{1-d}{2}\right)^3[\tanh^2(3m)+3\tanh^2(m) ]+d \left( \frac{1-d}{2} \right)^2 \\
\nonumber
&\times& 3\tanh^2(2m)+ 3 \frac{1-d}{2}d^2 \tanh^2(m)   \Big]\\
\nonumber
&\times& \Big[1-2 \left(\frac{1-d}{2} \right)^3[\tanh^2(3\beta m)+3\tanh^2(\beta m)]\\
&+&3d \left(\frac{1-d}{2}\right)^2 \tanh^2(2\beta m)+ 3\frac{1-d}{2}d^2\tanh^2(\beta m) \Big] \Big \}
\end{eqnarray}
has degeneracy $P-p$.

\begin{figure}\label{fig:simmetriciLD}
\begin{center}
\resizebox{0.60\columnwidth}{!}{\includegraphics{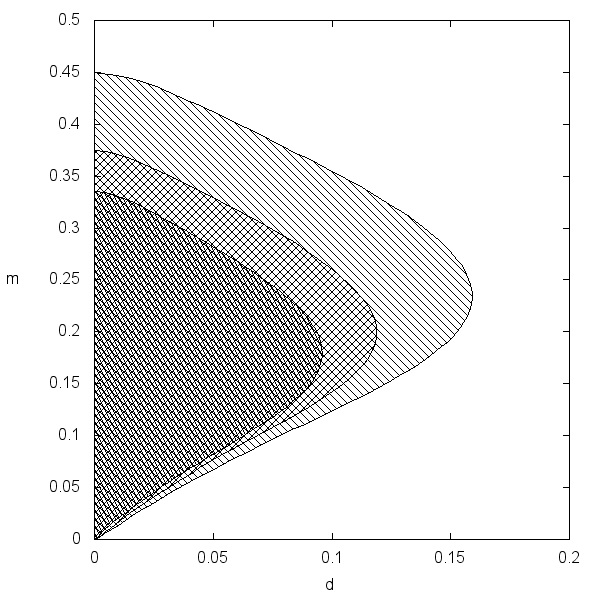}}
\caption{In this plot we focused on the region of the parameter space $(T,d)$, where odd symmetric spurious state exist and are stable. In particular, we chose $P=7$ and we considered any possible odd mixture, i.e. $p=3$, $p=5$ and $p=7$; each value of $p$ is represented by a different curve. Notice that the smaller $p$ and the wider the region, analogously to the standard Hopfield model.}
\end{center}
\end{figure}

Such states ($p<P$, $p$ odd) are stable only at small $d$.
%
This is due to the fact that the eigenvalue $c$ occurs only when $p<P$ and it reads as ($\mu>p$):
\begin{eqnarray}
\nonumber
A^{\mu\mu}&=&[1-\beta(1-d)]+\beta\langle (\xi^\mu)^2 \rangle_\xi \langle \tanh^2[\beta m \sum_\nu^p \xi^\nu] \rangle_\xi\\
&=&[1-\beta(1-d)]+\beta(1-d)\langle \tanh^2[\beta m \sum_\nu^p \xi^\nu] \rangle_\xi.
\end{eqnarray}
Thus, one can see that the r.h.s term contains factors $(1-d)$ at least of second order in such a way that when $d$ is close to $1$, i.e. for high dilution, and $T<1-d$, such term becomes negative. On the other hand, in the case $\mu \leq p$, we get
$$
A^{\mu\mu}=[1-\beta(1-d)]+\beta \langle  (\xi^\mu)^2 \tanh^2[\beta m \sum_{\nu=1}^p \xi^\nu] \rangle_\xi
$$
and therefore the r.h.s term contains even first order term $(1-d)$, which are comparable with $\beta(1-d)$.

Moreover, we find that the $p$-component, odd symmetric state exists and is stable in a region of the space $(T,d)$, which gets smaller and smaller as $p$ grows (see Fig.~ $5$). 
The emergence of such states  can be seen as a feature of robustness of the standard Hopfield model with respect to dilution.

Finally, the case $P=p$ always admits a region of existence and stability in the regime of high dilution. The latter region is independent of the parity and depends slightly on $P$ (see Fig.~$5$).
The emergence of such states is due to the failure of hierarchical retrieval, namely uniformity prevails.

\begin{figure}[!hd] \label{fig:simmetriciHD}
\begin{center}
\resizebox{0.60\columnwidth}{!}{\includegraphics{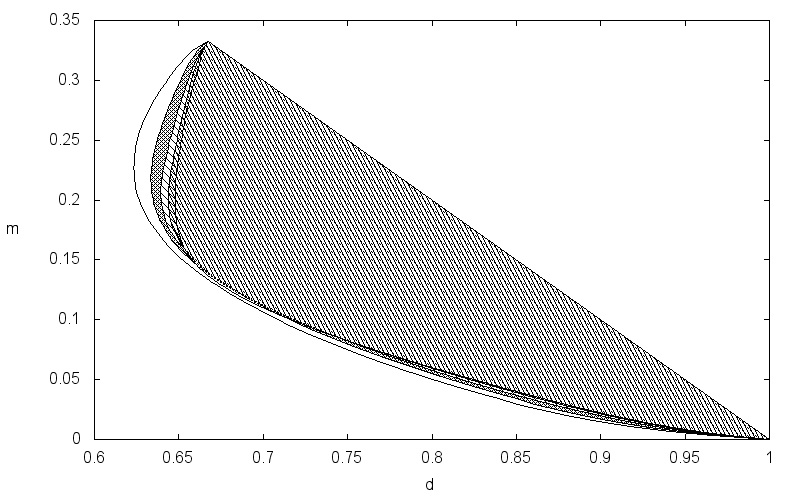}}
\caption{In this plot we focused on the region of the parameter space $(T,d)$, where  symmetric spurious state with $p=P$ exist and are stable. In particular, we chose $P=7$ and we considered any possible mixture, i.e. $p=3$, $p=4$, $p=5$, $p=6$ and $p=7$; each value of $p$ is represented by a different curve. Notice that the smaller $p$ and the wider the region, yet the region tends to an ``asymptotic shape''.}
\end{center}
\end{figure}


\subsection{Parallel state} \label{sec:parallel}
The parallel-retrieval state can be looked at as the extension to arbitrary values of $d$ of the pure state holding for the special case $T=0$.
We recall that in the noiseless limit the parallel-retrieval state can be described as
\be
\overrightarrow{m}=(1-d,(1-d)d,(1-d)d^2,...,(1-d)d^P).
\ee
In this case the stability matrix is diagonal with terms:
\be \label{eq:ps}
A^{\mu\mu}=1-\beta(1-d)+\beta\langle (\xi^\mu)^2\tanh^2[\beta(1-d)(\xi^1+d\xi^2+...+d^P \xi^P)]\rangle,
\ee
and, consistently, taking the limit $\beta \rightarrow \infty$, we get the simplified form
\be
A^{\mu\mu} = \lim_{\beta\rightarrow \infty}=1-\beta(1-d)+\beta\langle (\xi^\mu)^2(1-\delta[(\xi^1+d\xi^2+...+d^P \xi^P)])\rangle.
\ee
Now, the third term in the r.h.s. is either $\beta\langle (\xi^\mu)^2 \rangle = \beta (1-d)$ (when the polynomial of order $P$ is zero) or $0$; the latter case would trivially yield $A^{\mu\mu} <0$. Therefore, in the limit  $\beta \rightarrow \infty$ the stability of the parallel-retrieval state is constrained by the smallest real root $\in [0,1]$ of the polynomial $\xi^1 + d \xi^2 + ... + d^P \xi^P$ with $\xi^i=1,0,-1$. This corresponds to $\xi^1=1$ and $\xi^i=-1, \forall i>1$, under gauge symmetry and returns the same result found, from a more empirical point of view, in Sec.~\ref{sec:failure}. More precisely, the critical dilution converges exponentially to $1/2$ as $P$ grows.

In particular, for $P=3$ we find that the parallel-retrieval state exists and is stable in the interval $d\in(0,\frac{\sqrt{5}-1}{2})\simeq (0,0.618)$.
The point $d_c(3)=\frac{\sqrt{5}-1}{2}$ corresponds to the unique real root in $(0,1)$.

When noise is introduced, the critical dilution $d_c$, separating the parallel-retrieval state from spurious states, is shifted towards larger values, as suggested by Eq.~\ref{eq:ps}. On the opposite side, namely in the regime of small dilution, the parallel state is progressively depleted and, as the temperature is increased, magnetizations vanish, starting from $m_P$, and proceeding up to $m_2$. One can distinguish a set of temperatures $T_{P}(d) < T_{P-1}(d)<...<T_2(d)<T_1(d)$, such that when $T>T_K(d)$, all magnetizations $m_i, \forall i\leq K$ are null on average. Hence, above $T_2(d)$ the pure state retrieval is recovered, while above $T_1(d)=1-d$ the paramagnetic state emerges.


\begin{figure}[!hd] \label{fig:PR_P5}
\begin{center}
\resizebox{0.60\columnwidth}{!}{\includegraphics{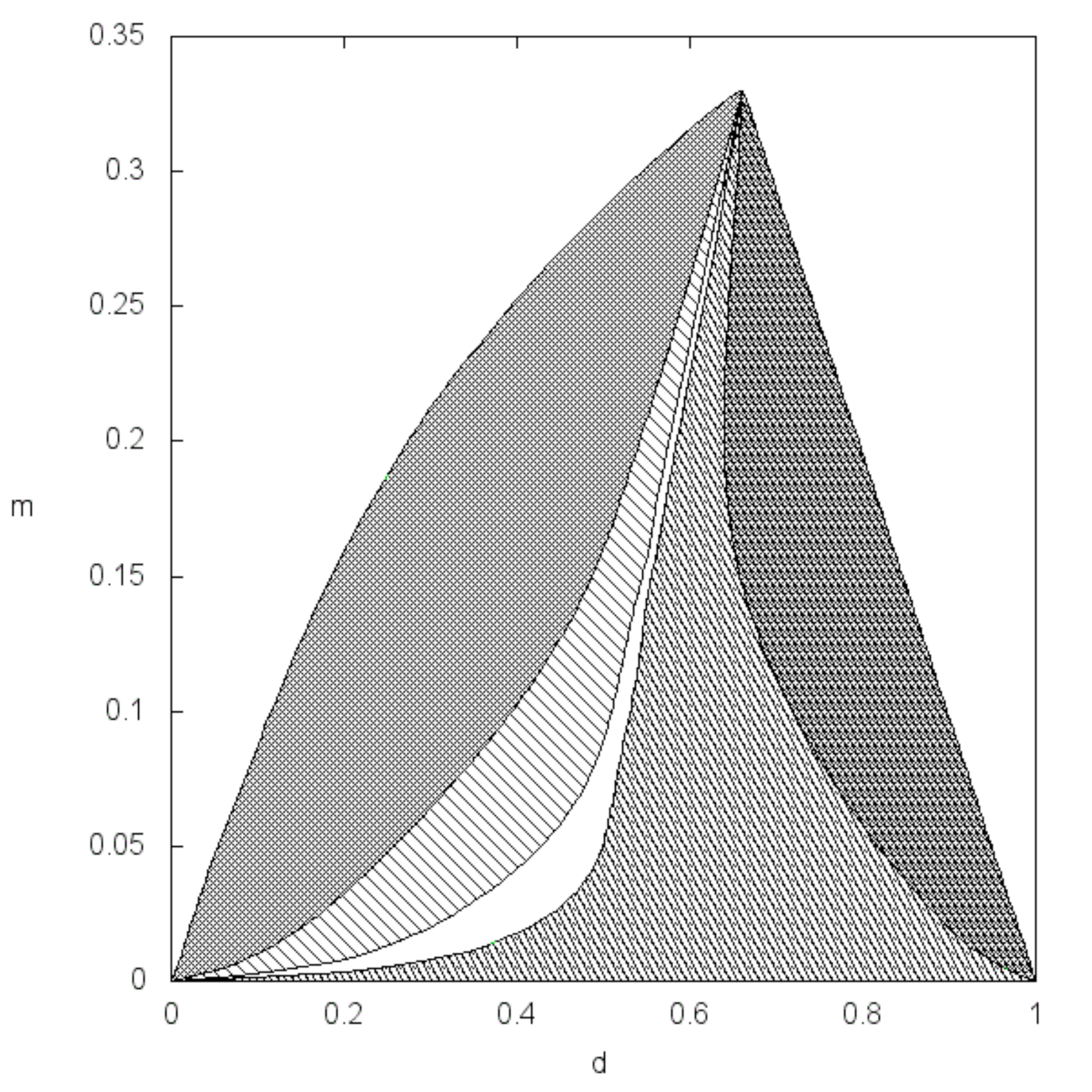}}
\caption{In this plot we focused on the region of the parameter space $(T,d)$, where parallel retrieval states exist and are stable. In particular, we chose $P=5$ and we considered any possible state with $k=2$, $k=3$, $k=4$ and $k=5$ non-null magnetization.}
\end{center}
\end{figure}

In Fig.~$7$  
we highlight the region of the parameter space $(T,d)$ where such parallel states exist and are stable. This was obtained numerically for the case $P=5$; for larger values of $P$ the region is slightly restricted to account for the shift in $d_c$.

\begin{figure}[h!]
 \begin{center}
\includegraphics[width=.3\textwidth]{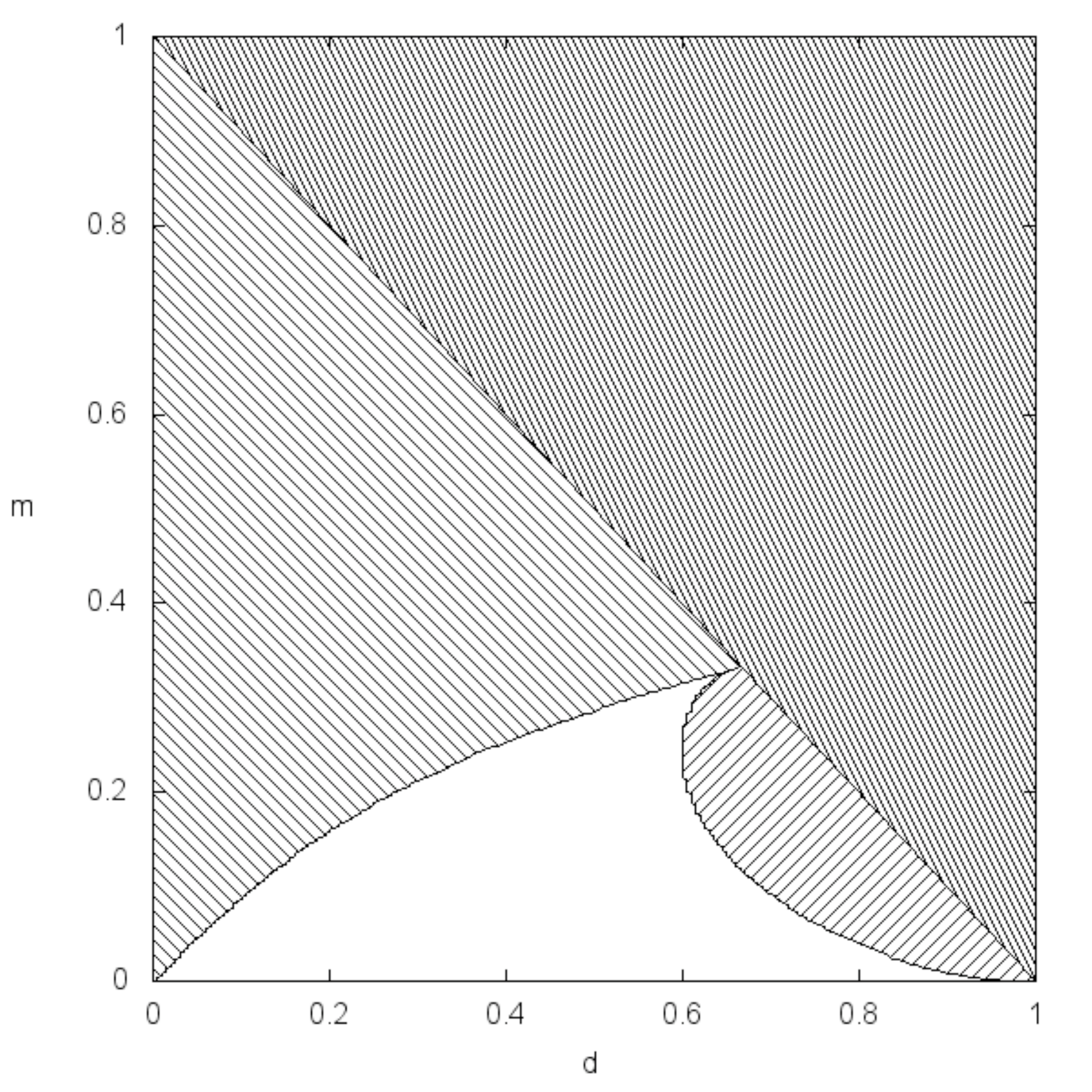}
\includegraphics[width=.3\textwidth]{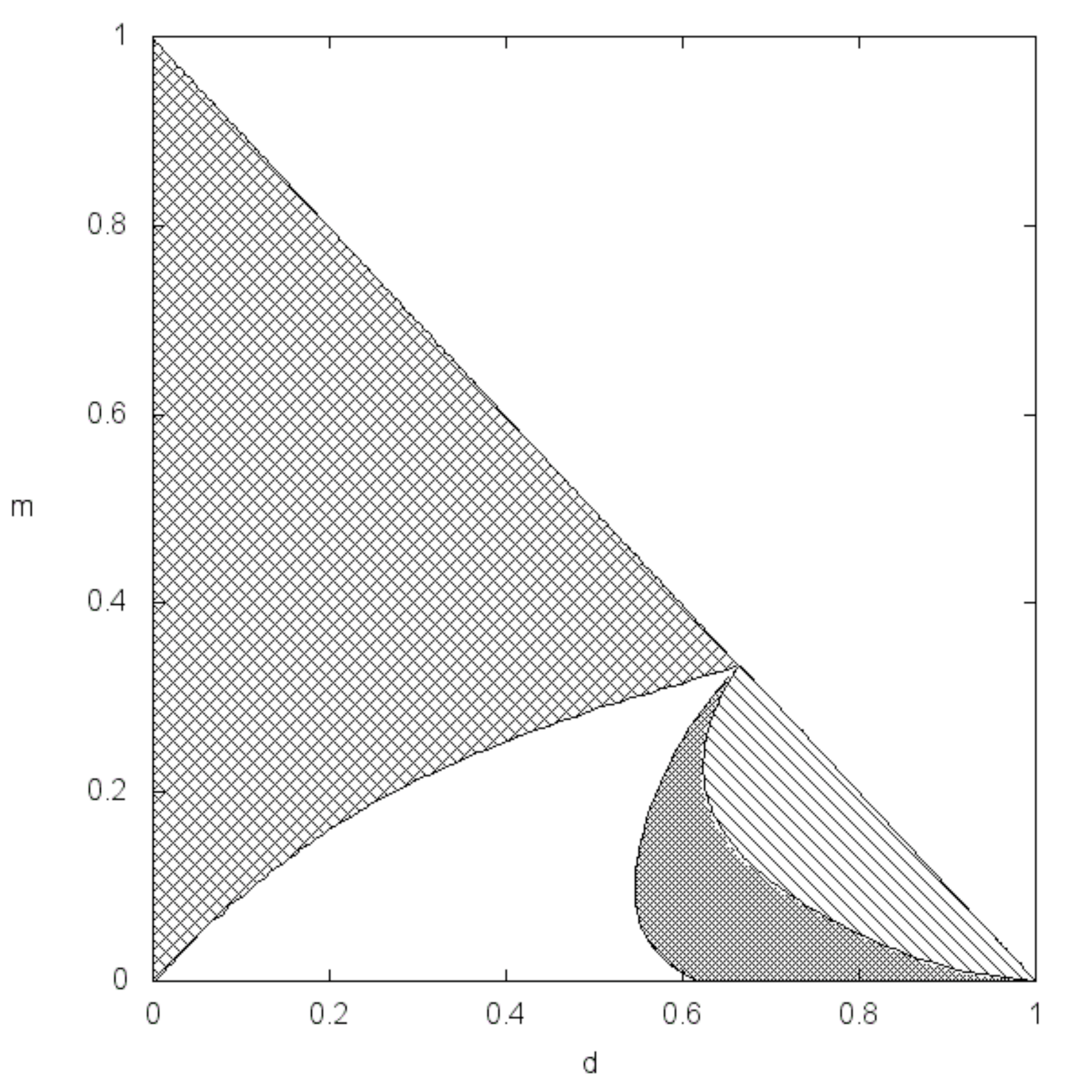}
\includegraphics[width=.3\textwidth]{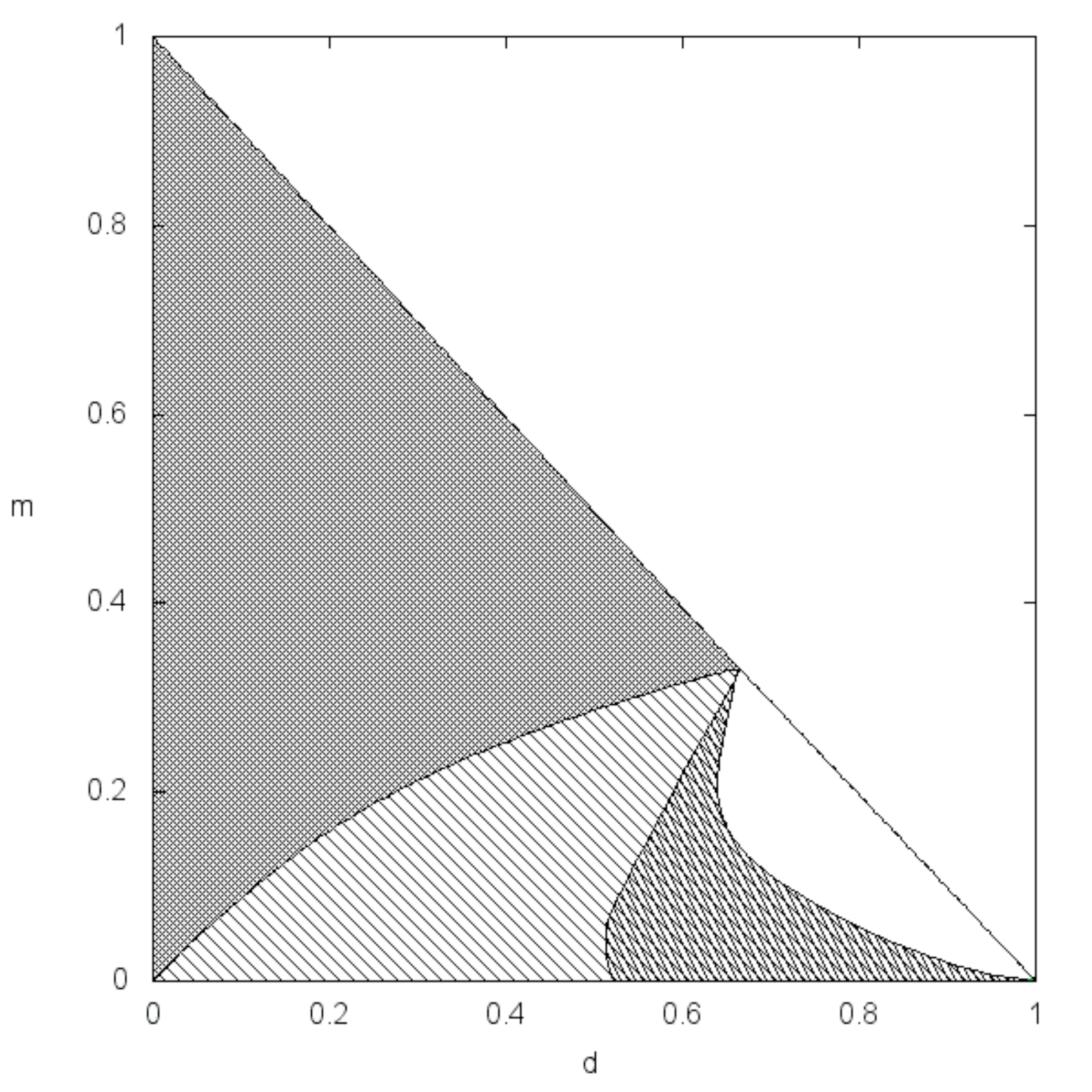}
\caption{\label{fig:PD} Phase diagram obtained from the analysis described in Sec.~\ref{sec:stability}. Each panel refers to a different value of $P$, namely $P=2$ (leftmost panel), $P=3$ (middle panel) and $P=5$ (rightmost panel). These theoretical predictions were also successfully compared with results from numerical simulations. Notice that when $P>3$, the region between the parallel states and the symmetric states includes spurious states, which are, in general a combination of the hybrid state and of the parallel state.
}
\end{center}
\end{figure}

Finally, the results collected so far are used to depict the phase diagrams for $P=2$, $P=3$ and $P=5$ (see Fig.~\ref{fig:PD}, from left to right).

\section{Discussion} \label{sec:discussion}
In this work we explored the retrieval capabilities of the multitasking associative network introduced in \cite{Agliari-2012PRL}.  Such a system is characterized by (quenched) patterns which display a fraction $d$ of null entries: interestingly, by paying the price of reducing the amount of information stored within each pattern (by  a fraction $d$), we get a system able to retrieve several patterns at the same time. Thus, this constitutes a model of a low information parallel processor; such a system can indeed be a good toy-model for all the phenomena where coordinated  multitasking features are expected as for instance in adaptive immune networks or peripheral nervous systems \cite{Agliari-sub,Londoner}.
\newline
At zero noise level ($T=0)$, and for a relatively low degrees of dilution, the system converges to an equilibrium state characterized by overlap $\mathbf{m} = ((1-d), (1-d)d, ..., (1-d)d^k, (1-d)d^{P-1})$, where $P$ is the number of stored patterns. Although this state displays non-null overlap with several patterns, it does not represent a spurious state, as can be seen by noticing, for instance, that this state allows the complete retrieval of at least one pattern. However, through a careful inspection, we proved in this paper that there are regions in the $(T, d)$ plane where genuine spurious states occur, hence a clear picture of the phase diagram becomes a fundamental issue in order to make the model ready for practical implementations.
\newline
A remarkable difference with respect to standard (serial processing) neural networks lies in the stability of mixture states: both even and odd mixtures are stable, which -within the world of spurious states - was a somewhat desired, and expected, result as there is neither a biological reason, nor a prescription from robotics, to weight differently odd and even mixtures (whose difference lies in the gauge invariance of the standard Hopfield model, which is broken within our framework due to the partial blankness of the pattern entries). Another expected feature, which we confirmed in this paper, is the emergence of parallel spurious states beyond standard ones from classical neural network theory: This is the natural generalization of the latter when moving from serial to parallel processing.

Beyond these somehow attended results, the phase diagram  of the model is still very rich and composed by several  not-overlapping regions where the retrieval states are deeply differently structured: Beyond the paramagnetic state and the pure state, the system is able to achieve both a hierarchical organization of pattern retrievals (for intermediate values of dilution) and a completely symmetric parallel state (for high values of dilution), which act as  the basis for the outlined mixtures when raising the noise level above thresholds  whose value depends on the load of the network $P$.
\newline
These findings have been obtained developing a new strategy for computing the free energy of the model by which, imposing thermodynamic principles (hence extremizing the latter over the order parameters of the model), self-consistency has been obtained.
The whole procedure has been strongly based on techniques stemmed from partial differential equation theory.
In particular, the key idea is showing that the noise-derivatives of the statistical pressure obey Burgers' equations, which can be solved through the Cole-Hopf transformation. The latter maps the evolution of the free energy over the noise into a diffusion problem which can be addressed through standard Green integration in momenta space and then pushed back in the original framework.
\newline
In the future, effort must still be spent in order to achieve a clear scenario in the hyper-diluted regime, namely where the dilution scales as a function of the volume (the amount of neurons), which can not be accomplished through the techniques we presented here as saddle point integration is no longer useful. We plan to report on this research soon.

%
%
%
%

\smallskip
\smallskip
\smallskip

\noindent
This work is supported by FIRB grant RBFR08EKEV. \\
Sapienza Universita' di Roma and INFN are acknowledged too for partial financial support. \\


\bibliography{listaANN2}

\begin{thebibliography}{10}

\bibitem{Amit-2003book}
D.~Amit.
\newblock {\em Modeling Brain Function}.
\newblock Cambridge University Press, 1989.

\bibitem{Amit-PRL1985}
D.J. Amit, H.~Gutfreund, and H.~Sompolinsky.
\newblock Storing infinite numbers of patterns in a spin-glass model of neural
  networks.
\newblock {\em Physical Review Letters}, 55:1530--1533, 1985.

\bibitem{Mezard-1987book}
M.~M\'ezard, G.~Parisi, and M.~A. Virasoro.
\newblock {\em Spin glass theory and beyond}.
\newblock World Scientific, Singapore, 1987.

\bibitem{Hopfield-1982PNAS}
J.J. Hopfield.
\newblock Neural networks and physical systems with emergent collective
  computational abilities.
\newblock {\em Proc. Natl. Acad. Sc. USA}, 79:2554--2558, 1982.

\bibitem{Cheng-NeuralNetworks1994}
B.~Cheng and D.~M. Titterington.
\newblock Neural networks: A review from a statistical perspective.
\newblock {\em Statistical Science}, 9(1):2--30, 1994.

\bibitem{Floreano-2000book}
S.~Nolfi and D.~Floreano.
\newblock {\em Evolutionary robotics: The biology, intelligence, and technology
  of self-organizing machines}.
\newblock 2000.

\bibitem{Trippi-1992book}
R.R. Trippi and E.~Turban.
\newblock {\em Neural Networks in Finance and Investing: Using Artificial
  Intelligence to Improve Real World Performance}.
\newblock McGraw-Hill, Inc. New York, NY, USA, 1992.

\bibitem{Agliari-JTB2011}
E.~Agliari, A.~Barra, F.~Guerra, and F.~Moauro.
\newblock A thermodynamic perspective of immune capabilities.
\newblock {\em J. Theor. Biol.}, 287:48--63, 2011.

\bibitem{Sompolinsky-1986PRA}
H.~Sompolinsky.
\newblock Neural networks with non-linear synapses and a static noise.
\newblock {\em Physical Review A}, 34:2571, 1986.

\bibitem{Coolen-2004JPA}
T.~Nikoletopoulos, A.C.C. Coolen, I.~P{\'e}rez-Castillo, N.S. Skantzos, J.P.L.
  Hatchett, and B~Wemmenhove.
\newblock Replicated transfer matrix analysis of ising spin models on 'small
  world' lattices.
\newblock {\em Journal of Physics A: Mathematical and General}, 37:6455, 2004.

\bibitem{Coolen-2003JPA}
B.~Wemmenhove and A.C.C. Coolen.
\newblock Finite connectivity attractor neural networks.
\newblock {\em Journal of Physics A Mathematical and Theoretical}, 36(9617),
  2003.

\bibitem{Barra-2010JSP}
A.~Barra, F.~Guerra, and G.~Genovese.
\newblock The replica symmetric approximation of the analogical neural network.
\newblock {\em Journal of Statistical Physics}, 140(4):784, 2010.

\bibitem{Barra-2012NN}
A.~Barra, A.~Bernacchia, E.~Santucci, and P.~Contucci.
\newblock On the equivalence of hopfield networks and boltzmann machines.
\newblock {\em Neural Networks}, 2012.

\bibitem{Agliari-2012PRL}
E.~Agliari, A.~Barra, A.~Galluzzi, F.~Guerra, and F.~Moauro.
\newblock Multitasking {A}ssociative {N}etworks.
\newblock {\em Physical Review Letters}, 109:268101, 2012.

\bibitem{Agliari-2012NN}
E.~Agliari, A.~Barra, A.~De Antoni, and A.~Galluzzi.
\newblock Parallel retrieval of correlated patterns: {F}rom {H}opfield networks
  to {B}oltzmann machines.
\newblock {\em Neural Networks}, 38:52--63, 2012.

\bibitem{Coolen-2005book}
A.C.C. Coolen, R.~K\"{u}hn, and P.~Sollich.
\newblock {\em Theory of Neural Information Processing Systems}.
\newblock Oxford University Press, 2005.

\bibitem{Agliari-sub}
E.~Agliari, A.~Barra, S.~Bartolucci, A.~Galluzzi, F.~Guerra, and F.~Moauro.
\newblock Parallel processing in immune networks.
\newblock {\em submitted}, 2012.

\bibitem{Willshaw-1976PRSL}
D.J. Willshaw and C.~von~der Malsburg.
\newblock How patterned neural connections can be set up by self-organization.
\newblock {\em Proc. R. Soc. Lond. B}, 194:431--445, 1976.

\bibitem{Barra-2010JSM}
A.~Barra, A.~Di~Biasio, and F.~Guerra.
\newblock Replica symmetry breaking in mean-field spin glasses through the
  hamilton--jacobi technique.
\newblock {\em Journal of Statistical Mechanics: Theory and Experiment},
  2010:P09006, 2010.

\bibitem{Genovese-2009JMP}
G.~Genovese and A.~Barra.
\newblock A mechanical approach to mean field models.
\newblock {\em Journal of Mathematical Physics}, 50:053303, 2009.

\bibitem{Londoner}
E.~Agliari, A.~Annibale, A.~Barra, A.C.C. Coolen, and D.~Tantari.
\newblock Immune networks: {M}ultitasking properties at medium load.
\newblock {\em submitted}, 2013.

\end{thebibliography}
\bibliographystyle{unsrt}

\end{document}